# Sum-Rate Optimal Power Policies for Energy Harvesting Transmitters in an Interference Channel

Kaya Tutuncuoglu  Aylin Yener
Wireless Communications and Networking Laboratory (WCAN)
Electrical Engineering Department
The Pennsylvania State University, University Park, PA 16802
kaya@psu.edu    yener@ee.psu.edu

*Abstract*—This paper considers a two-user Gaussian interference channel with energy harvesting transmitters. Different than conventional battery powered wireless nodes, energy harvesting transmitters have to adapt transmission to availability of energy at a particular instant. In this setting, the optimal power allocation problem to maximize the sum throughput with a given deadline is formulated. The convergence of the proposed iterative coordinate descent method for the problem is proved and the short-term throughput maximizing offline power allocation policy is found. Examples for interference regions with known sum capacities are given with directional water-filling interpretations. Next, stochastic data arrivals are addressed. Finally online and/or distributed near-optimal policies are proposed. Performance of the proposed algorithms are demonstrated through simulations.

## I. INTRODUCTION

Recent advances in energy harvesting and the advocacy for green technologies are leading to significant interest in systems powered by harvested ambient energy. In wireless communications, energy harvesting does more than reducing the carbon footprint of today's high rate wireless voice and data systems: It also makes available self-sustaining wireless networks with an indefinite lifetime. Such ease of deployment and maintenance for wireless nodes as well as a growing demand for high rate communications foresee a rapid increase in number of energy harvesting communication devices in the near future. Given these advantages, one can forecast that there shall be growing interest in wireless networks comprised of energy harvesting nodes.

The design principles of energy harvesting wireless networks are fundamentally different than their traditional counterparts: In order to utilize the wireless and the energy resources in the best possible way, the network needs to be optimized subject to the constraints on the *instantaneously available* energy. Energy availability is stochastic and uneven throughout operation, and the battery to store the harvested energy is limited in practice. A particularly important network structure is one that addresses the case where multiple energy harvesting transmitters share the wireless medium to communicate to multiple destinations, i.e., a wireless ad hoc network with interference. In this paper, we consider the simplest such setting, the two-user interference channel, and solve the optimum power scheduling problem that maximizes short-term sum throughput of this system under a deadline when the two transmitters obtain the transmission energy by harvesting from ambient sources.

Optimal power policies for energy harvesting nodes have attracted recent interest in the research community. As mentioned above, the essence of the problem is adapting transmission power to energy availability. One approach is to sustain a performance while preserving a balance between harvested and consumed energy, i.e., energy neutrality [1]. References [2], [3] provide energy neutrality by stabilizing an energy queue. This approach often calls for the energy queue to grow indefinitely large for optimal operation, thus is not applicable to nodes with limited battery capacity. An alternative approach enforcing strict energy constraints is considered in [4], where transmission time for a given amount of data is optimized over power allocations that obey a known energy arrival scheme. This work has been subsequently extended to the problem of maximizing transmitted data by a deadline with the addition of a battery capacity constraint in [5], and battery imperfections in [6]. A model incorporating channel fading to these problems is introduced in [7] and solved using a directional water-filling algorithm. While these early works considered a single user setting, i.e., one energy harvesting transmitter and one receiver, more recently, settings with either one energy harvesting transmitter broadcasting to multiple receivers [8], [9], or multiple energy-harvesting transmitters communicating to one receiver [10] have also been considered. In contrast, we, in this paper, focus on the scenario with multiple transmitters and multiple receivers, i.e., the interference channel.

The interference channel is a fundamental building block for wireless networks. Consequently, identifying the "correct" transmission policies under energy harvesting scenario for this channel, will furnish us with insights needed for energy harvesting wireless ad hoc network design. A critical issue is the lack of conclusive results on the capacity of the interference channel. For the Gaussian two-user interference channel, which is our focus, the strong interference capacity region was characterized earlier in [11]. Additional recent results with respect to the capacity region and weak interference sum capacity have been obtained in [12], [13]. The known sum-capacity results point out the fact that the capacity is notably influenced by the interaction of the transmitters, and how

This work was supported by NSF Grant CNS 0964364.

interference is processed at the receivers [12]. Considering that the energy availabilities of energy harvesting nodes are varying, the problem of optimal power allocation in this setting becomes an interesting one to tackle.

The focus of this paper is on short-term throughout optimization in a two-user Gaussian interference channel with energy harvesting transmitters. The problem of transmitting the maximum total number of bits for a given deadline is considered. First, it is shown that an iterative coordinate descent algorithm optimizing individual power allocations at each iteration converges to the optimal solution for a jointly concave sum-rate expression when all data is available for transmission beforehand. This suggests performing single user generalized directional water-filling algorithms iteratively alternating between the users to find the optimal power allocation. Then, examples for specific interference regions are presented for which parts of the solution reduces to simpler directional water-filling algorithms. In general, it is observed that variations of the directional water-filling algorithm is necessary to adapt to energy arrivals and interference in the optimal policy. Next, the solution is extended to the scenario when data arrivals, just like harvested energy, occur intermittently during the communication. A modified version of the directional water-filling algorithm which handles data causality through a penalty function was proposed. Finally, distributed and online algorithms are proposed.

The paper is structured as follows. In Section II, we describe the general system model, with energy and packet arrivals to both transmitters in any interference region as well as the problem definition. In Section III, we present the iterative algorithm and prove its convergence. Some examples for the iterative algorithm in different interference regions are given in Section IV with all data available at the beginning of transmission. In Section V, the extension to a setting with intermittent packet arrivals is considered. In Section VI, some near-optimal algorithms are proposed, which are subsequently simulated in Section VII. Section VIII concludes the paper.

## II. SYSTEM MODEL

The two-user Gaussian interference channel with energy harvesting transmitters is shown in Figure 1. Transmitters $T1$ and $T2$ have independent data packets addressed to corresponding receivers $R1$ and $R2$. The transmitters are powered by independent energy harvesting processes, the energy from which are stored in batteries of size $E_{1,max}$ and $E_{2,max}$ respectively. The harvested energies and battery capacities are normalized to the corresponding transmitter-receiver link gain and receiver noise level, yielding unitary direct channel coefficients and noise variances. After this normalization for each transmitter, the cross channel coefficients become $\sqrt{a}$ and $\sqrt{b}$ and the channel outputs are expressed as

$$Y_1 = X_1 + \sqrt{a}X_2 + Z_1, \qquad Y_2 = \sqrt{b}X_1 + X_2 + Z_1 \quad (1)$$

where $Y_1$ and $Y_2$ are received at $R1$ and $R2$, $X_1$ and $X_2$ are channel inputs by $T1$ and $T2$ normalized to have unit channel gains at their corresponding receivers, and $Z_1$ and $Z_2$ are zero-mean Gaussian random variables with unit variance.

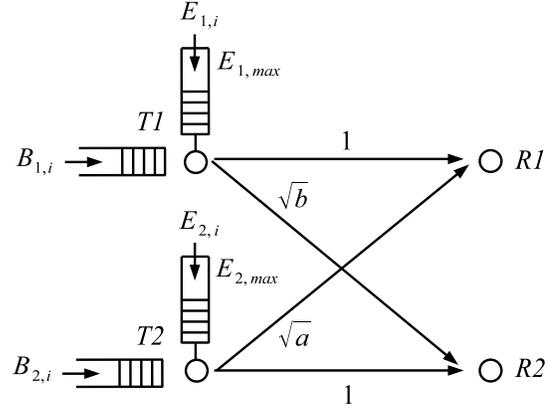

Fig. 1: Interference channel model with energy harvesting and data arrivals.

Using this model, reference [12] reports sum capacity results for the Gaussian interference channel for ranges for $a$ and $b$ as summarized in [12, Table 1]. The sum rate as a function of the powers of transmitters will be referred to as the power-rate function $r(p_1, p_2)$ in the sequel, and individual achievable rates will be denoted with $r_j(p_1, p_2)$. Specific regions for $a$ and $b$ will be denoted as a superscript when needed. It is assumed for the sake of simplicity that transmission is the dominant source of energy expenditure in the system, and other factors such as base power or processing power are ignored. Possible effects of these factors are discussed in Section III.

The energy harvesting process and the packet arrival process for node $j \in 1, 2$ are denoted in Figure 2 with red and blue arrows respectively. We assume a time slotted system[1] with slots of length $\tau$, where a normalized energy of $E_{j,i}$ units and a data packet of size $B_{j,i}$ bits are received by transmitter $j$ at the beginning of time slot $i$ and is available for immediate use within that time slot. Since an instantaneous energy consumption requires infinite instantaneous power which is impractical, the energy harvests must be stored in the battery before consumption. Thus any arrival exceeding the respective battery capacity is irreversibly lost, and an arrival larger than the respective battery capacity is truncated in the model accordingly. Arriving data packets are stored in the data buffer as well, only without a buffer size restriction. For optimal policy analysis, it is assumed in Sections III-V that the arrival scheme is perfectly and non-causally known by both transmitters before transmission. This problem is referred to as the *offline* problem. Near-optimal algorithms with online or non-centralized decisions are put forth in Section VI.

There are multiple constraints in this model for a feasible selection of a transmission policy. The first constraint is the energy causality in the sense that no more than the already harvested amount of energy shall be consumed up to a time in transmission. Denoting the transmission power of user $j$ over time slot $i$ as $p_{j,i}$, the constraint for time slot $n$ can be

---

[1]A time-slotted model is preferred over the continuous time discrete arrival models in [4], [5], [7] for notational simplicity.

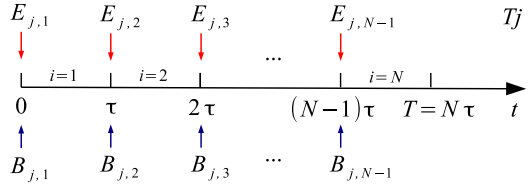

Fig. 2: Energy harvests and data arrivals in the time-slotted model.

expressed as
$$\sum_{i=1}^{n} E_{j,i} - \sum_{i=1}^{n} \tau \cdot p_{j,i} \geq 0 \quad (2)$$

where $j$ is the transmitter index chosen from the set $\{1,2\}$. Note that assuming constant power transmission within a time slot is optimal, as proved in [4]. Secondly, it is shown in [5] that a battery overflow is undesirable since any overflowing energy can be consumed prior to the overflow, strictly increasing the utility. This argument applies to the interference channel as well since $r(p_1, p_2)$ is increasing in $p_1$ and $p_2$ by definition. The implication is that at the end of time slot $n$, there should be sufficient space in the battery to accommodate the next harvest $E_{n+1}$. Therefore the battery capacity constraint,
$$\sum_{i=1}^{n} \tau \cdot p_{j,i} + E_{j,max} - \sum_{i=1}^{n+1} E_{j,i} \geq 0 \quad (3)$$

is to be met for every $n$ over the transmission. Note that it is possible for a transmitter to not have any extra bits in the data queue when a battery overflow is imminent, rendering overflow avoidance practically useless. This special case is discussed in detail in Section V. The final constraint is data causality, implying that no more than the available amount of data can be transmitted until the end of the $n^{th}$ slot for every $n$ throughout the transmission,
$$\sum_{i=1}^{n} B_{j,i} - \sum_{i=1}^{n} \tau \cdot r_j(p_{j,i}) \geq 0. \quad (4)$$

We define the problem of maximizing the total number of bits sent by the transmitter until a deadline $T = N \cdot \tau$, i.e., $N$ time slots, as the *short-term throughput maximization* problem, which can be expressed as follows:
$$\max_{\mathbf{p_1} \geq 0, \mathbf{p_2} \geq 0} \sum_{i=1}^{N} \tau \cdot r(p_{1,i}, p_{2,i}) \quad (5a)$$
$$\text{s.t.} \sum_{i=1}^{n} (E_{j,i} - \tau \cdot p_{j,i}) \geq 0, \quad (5b)$$
$$\sum_{i=1}^{n} (\tau \cdot p_{j,i} - E_{j,i}) + E_{j,max} - E_{j,i+1} \geq 0, \quad (5c)$$
$$\sum_{i=1}^{n} (B_{j,i} - \tau \cdot r_j(p_{j,i})) \geq 0, \quad n=1,...,N, \ j \in \{1,2\}. \quad (5d)$$

In (5), the vector $\mathbf{p_j}$ represents the collection of transmission powers of user $j$, and will be referred to as the power policy or the power allocation vector of user $j$ in the sequel. Constraints (5b) and (5d) correspond to energy causality, battery constraint and data causality respectively. The expression $\mathbf{p_j} \geq 0$ implies component-wise non-negativity on the transmission power vector.

We shall first focus on the case where an infinite backlog of data is available at both transmitters at the beginning of transmission, and ignore the data causality constraint given in (5d). We include (5d), and extend our approach to the problem with data arrivals in Section V.

## III. ITERATIVE SOLUTION

In this section, we employ an iterative approach to solve the two user optimization problem defined in (5) without the data causality constraint in (5d). In particular, we show the convergence of the cyclic coordinate descent method where the two coordinates are chosen as the power allocation vectors of the two users, namely $\mathbf{p_1}$ and $\mathbf{p_2}$.

### A. Iterative Algorithm

The proposed algorithm is to solve the two user interference channel problem by iteratively maximizing the throughput over the transmission policy of one user while keeping the other policy constant until both policies converge to the optimal vector. Starting from $T1$ and an arbitrary initial feasible pair $(\mathbf{p_1^0}, \mathbf{p_2^0})$, the following update for the power policies is performed on the $k^{th}$ iteration:
$$\mathbf{p_1^k} = \arg\max_{\mathbf{p_1} \geq 0} \sum_{i=1}^{N} \tau \cdot r(p_{1,i}, p_{2,i}^{k-1})$$
$$\text{s.t.} \sum_{i=1}^{n} (\tau \cdot p_{1,i} - E_{1,i}) + E_{1,max} - E_{1,n+1} \geq 0,$$
$$\sum_{i=1}^{n} (E_{1,i} - \tau \cdot p_{1,i}) \geq 0, \quad n=1,...,N \quad (6)$$
$$\mathbf{p_2^k} = \arg\max_{\mathbf{p_2} \geq 0} \sum_{i=1}^{N} \tau \cdot r(p_{1,i}^k, p_{2,i})$$
$$\text{s.t.} \sum_{i=1}^{n} (\tau \cdot p_{2,i} - E_{2,i}) + E_{2,max} - E_{2,n+1} \geq 0,$$
$$\sum_{i=1}^{n} (E_{2,i} - \tau \cdot p_{2,i}) \geq 0, \quad n=1,...,N \quad (7)$$

Note that the energy constraints for only the optimized vector is included in each problem. This is due to the other vector being fixed as the output of a previous iteration, or as the feasible initial point, implying its feasibility regardless of the value of the optimized vector. This is possible since the constraints are not coupled due to the transmitters harvesting energy independently and consuming their own energy. The problems in (6) and (7) involve single-user optimization of a sum of concave functions over a linear set of constraints, a similar of which was solved in [14] utilizing a generalized water-filling algorithm. In order to conform to energy causality and battery capacity constraints, this algorithm needs to be enhanced as in [7] with directional water-flow and taps, as demonstrated by the examples in Section V.

### B. Convergence

Optimality of the iterative solution follows from the convexity of the problem and the constraints. We start by stating that the rate function in the objective of (5) as well as (6) and (7) can be considered concave and non-decreasing without loss of generality.

*Lemma 1:* Given any coding scheme achieving an instantaneous rate of $r'(p_1, p_2)$, one can construct a scheme achieving a rate $r(p_1, p_2)$ concave and non-decreasing in $p_1$ and $p_2$, that performs no worse than the given scheme.

*Proof:* The proof for the non-decreasing property is straightforward, since a scheme can always choose to discard some of the allocated power to achieve the rate of a smaller power vector. The concavity property is shown using the following time-sharing argument. Given $r'(p_1, p_2)$ for $p_1, p_2 \geq 0$, define

$$r(p_1, p_2) = \max \left\{ \begin{array}{l} \sum_i \lambda_i r'(p_1^{(i)}, p_2^{(i)}) \quad s.t. \quad \lambda_i \geq 0, \\ \sum_i \lambda_i = 1, \quad \sum_i \lambda_i p_j^{(i)} = p_j, \quad j = 1, 2 \end{array} \right\} \quad (8)$$

as the maximum achievable rate using the given scheme and time-sharing. Note that all rates in the max term in (8) are achievable by time-sharing between the points $(p_1^{(i)}, p_2^{(i)})$ with corresponding parameters $\lambda_i$, while consuming the same average power for both users. For all power vectors, $r(p_1, p_2)$ is at least as good as $r'(p_1, p_2)$ since $r'(p_1, p_2)$ can be achieved with $\lambda_1 = 1$. Finally, any rate achieved will be a convex combination of a number of points on $r'(p_1, p_2)$, which means that a convex combination of two points can also be expressed as such. Since the rate at any point is the maximum of such convex combinations, $r(p_1, p_2)$ is also jointly concave in $p_1$ and $p_2$. Thus, $r(p_1, p_2)$ defined in (8) is an achievable, concave scheme that performs at least as good as $r'(p_1, p_2)$ everywhere. ∎

We note that implementing such a scheme is feasible when the time scale considered is sufficiently long to allow time-sharing. Thus the actual capacity of this channel will also have to be concave, as is the case for all known interference channel regions with known sum-capacities.

With the derivation of a concave rate function that we can effectively replace any coding scheme with, we state the convergence of the iterative algorithm.

*Theorem 1:* The iterative algorithm given in (6) and (7) converges to the optimal policy.

*Proof:* The iterative optimization approach among the variables of the problem, commonly referred to as the *block coordinate descent* method, is known to converge for a problem in the form of

$$\max f(x_1, x_2, ..., x_n) \quad s.t. \quad \boldsymbol{x} \in \mathcal{X} \quad (9)$$

when the objective function $f$ is continuously differentiable over $\mathcal{X}$, and the feasible set $\mathcal{X}$ can be expressed as the Cartesian product of convex sets $X_1, ..., X_n$. Furthermore, it is required for the objective function to yield a unique maximum in all variables $x_i$, i.e.,

$$\max_{\zeta \in X_i} f(x_1, x_2, ..., x_{i-1}, \zeta, x_{i+1}, ..., x_j) \quad (10)$$

needs to have a unique $\zeta$ solving this problem [15, Prop. 2.7.1].

In the throughput maximization problem, we propose to perform the iterations over the power policies of the two users, partitioning the variable space into two, namely $(\boldsymbol{p_1}, \boldsymbol{p_2})$, yielding the iterations in (6,7). Since the two nodes harvest and consume the energy independently, the set of constraints on $\boldsymbol{p_1}$ and $\boldsymbol{p_2}$ can be separated. The two constraint sets are also convex, since the individual constraints are linear in their respective elements $p_{j,i}$. Thus the constraint sets do satisfy the requirements for convergence.

For the objective function, we assume that $r(\boldsymbol{p_1}, \boldsymbol{p_2})$ is continuously differentiable, which would hold for any well-behaved achievable scheme over which $r(\boldsymbol{p_1}, \boldsymbol{p_2})$ is constructed, as is the case for all known interference channel sum-capacity expressions. Additionally, it is required that the property in (10) is satisfied, which is yielding a unique maximum for either user when the policy for the other user is kept constant. This requirement is trivially satisfied for a strictly concave objective function on a convex set $X_i$. However, as stated by Lemma 1, for an interference setting without an explicit capacity definition, one can only guarantee concavity, and any claim of strict concavity is violated whenever time-sharing is used. This is overcome by introducing two auxiliary vectors $\mathbf{s_1}$ and $\mathbf{s_2}$ and restating the maximization problem with the objective function

$$g(\mathbf{p_1}, \mathbf{p_2}, \mathbf{s_1}, \mathbf{s_2}) = r(\mathbf{p_1}, \mathbf{p_2}) - \epsilon \|\mathbf{p_1} - \mathbf{s_1}\|^2 - \epsilon \|\mathbf{p_2} - \mathbf{s_2}\|^2 \quad (11)$$

to replace $r(\mathbf{p_1}, \mathbf{p_2})$ in (5), where $\epsilon > 0$ is an arbitrary coefficient. It can be observed that the modified objective $g(\mathbf{p_1}, \mathbf{p_2}, \mathbf{s_1}, \mathbf{s_2})$ is strictly concave for a concave $r(\mathbf{p_1}, \mathbf{p_2})$, and thus a convergence statement holds for an algorithm using this objective for any positive $\epsilon$. Such an algorithm would cycle between $\mathbf{p_1}, \mathbf{p_2}, \mathbf{s_1}$ and $\mathbf{s_2}$ in its iterations. The solutions to the iterations of auxiliary variables are trivial, with $\mathbf{s_1}$ and $\mathbf{s_2}$ assuming the values of $\mathbf{p_1}$ and $\mathbf{p_2}$ respectively to minimize the euclidian distances in (11). On the other hand, iteration steps on $\mathbf{p_1}$ and $\mathbf{p_2}$ suffer from the additional terms in (11) that increases as they move away from $\mathbf{s_1}$ and $\mathbf{s_2}$, which are equal to their values in the previous iteration. In essence, this modification employs a penalty on moving away from the previous value of the power policy, and can be thought of choosing the closest maximizer when the maximum is not unique if the values of $\epsilon$ is taken to be sufficiently small. Since the block coordinate descent method for this strictly concave cost function converges for an arbitrarily small $\epsilon$ [15], so does the proposed iterative algorithm in (6) and (7) provided that in case of multiple maximizers, the one closer to the previous power vector is favored. We shall refer to this requirement as the *minimum displacement rule* in the sequel. ∎

## IV. EXAMPLES

With the convergence of the iterative algorithm verified, we now present examples on how the algorithm is applied in interference channels with known sum capacities. For convenience, we briefly summarize the directional water-filling algorithm introduced in [7] since it provides valuable insight on the solution of the single user subproblems of the iterative algorithm.

### A. Directional Water-Filling for a Fading Channel

Consider a fading communication channel with power-rate function $r(p) = \frac{1}{2}log(1 + h \cdot p)$ and an energy harvesting transmitter with energy constraints (2) and (3) as well as $p_k > 0$ for $k = 1, \ldots, N$. To find the optimal power policy for the short term throughput maximization problem, we first compute its Lagrangian as

$$\mathcal{L} = \tau \sum_{i=1}^{N} \frac{1}{2}log(1 + h_i \cdot p_i) - \sum_{k=1}^{N} \lambda_k \left(\sum_{i=1}^{k} \tau \cdot p_i - E_i\right) \quad (12)$$
$$- \sum_{k=1}^{N-1} \mu_k \left(\sum_{i=1}^{k}(E_i - \tau \cdot p_i) + E_{k+1} - E_{max}\right) - \sum_{k=1}^{N-1} \eta_k \cdot p_k$$

with Lagrangian multipliers $\{\lambda_i\}$, $\{\mu_i\}$ and $\{\eta_i\}$. Applying the Karush-Kuhn-Tucker (KKT) stationarity condition by evaluating the gradient of $\mathcal{L}$ gives the optimal power policy as

$$p_i^* = \left[\nu_i - \frac{1}{h_i}\right]^+, \qquad \nu_i = \frac{1}{\sum_{k=i}^{N}(\lambda_k - \mu_k)} \quad (13)$$

where $\nu_i$ is the water level at the $i^{th}$ time slot, evaluated as shown above. The KKT complementary slackness conditions indicate that $\lambda_i$ and $\mu_i$ are positive only when the battery is empty and full respectively, and zero otherwise; implying that water level increases and deceases only when the corresponding constraint is satisfied with equality. This yields a directional water-filling interpretation with base level $1/h_i$, that satisfies energy causality by allowing water flow in only forward direction, and battery capacity by not allowing more than $E_{max}$ amount of water to flow between any two time slots. The flow constraints ensure that increasing and decreasing water levels only occur when the corresponding constraint is active, and water level is equalized elsewhere, thus solving the power allocation problem for a single user in a Gaussian fading channel [7].

### B. Asymmetric Interference with $ab > 1$

Asymmetric interference region is where one transmitter has a strong cross channel gain and the other has a weak one. It covers the two symmetric cases $a \leq 1$, $b \geq 1$ and $a \geq 1$, $b \leq 1$. We shall assume the former case in this paper, with the result easily applicable to the latter case by switching transmitter indices.

The capacity achieving scheme in the asymmetric interference with $ab > 1$ is treating the weaker interference as noise and decoding and removing the stronger interference at the receiver. The power-rate function for this region and is given as [12]

$$r_A(p_1, p_2) = \frac{1}{2}log\left(1 + \frac{p_1}{1 + ap_2}\right) + \frac{1}{2}log(1 + p_2). \quad (14)$$

We consider the iterations for each user separately while assuming that the power allocation policy of the other user is fixed. Denoting the sum rate as a function of only user $j$'s power as $r(p_1)$, the single user power-rate function for $T1$ can be expressed as

$$r_A(p_1) = \frac{1}{2}log(1 + h \cdot p_1) + C_1 \quad (15)$$

with $h = \frac{1}{1+ap_2}$ and $C_1 = \frac{1}{2}log(1+p_2)$. The direct implication of this form is treating the interference term as channel fading $h$ and using the directional water-filling algorithm. The solution for $T1$ reduces to directional water-filling in Subsection IV-A, for which the channel fading parameter $h$ in each time slot are updated with the output of the previous iteration for the second user, $\boldsymbol{p}_2^{k-1}$ at the $k^{th}$ iteration.

For $T2$, the single user power-rate function remains as the sum of two terms involving $p_2$ and therefore does not simplify to a common form. Instead, we tackle the problem by evaluating the KKT optimality conditions. The stationarity condition requires

$$\frac{d}{dp_2}r_A(p_2)\bigg|_{p_{2,n}} - \sum_{i=n}^{N}(\lambda_{2,i} - \mu_{2,i}) - \eta_{2,n} = 0 \quad (16)$$

on the $n^{th}$ time slot for every $0 \leq n \leq N$, and $\eta_{2,n}$ is the Lagrange multiplier corresponding to the nonnegativity constraint on the power level, $\boldsymbol{p}_2 \geq 0$. The complementary slackness conditions for this problem arise as

$$\lambda_{2,n}\left(\sum_{i=1}^{n} E_{2,i} - \sum_{i=1}^{n} \tau \cdot p_{2,i}\right) = 0, \quad \lambda_{2,n} \geq 0,$$
$$\mu_{2,n}\left(\sum_{i=1}^{n} \tau \cdot p_{2,i} + E_{2,max} - \sum_{i=1}^{n+1} E_{2,i}\right) = 0, \quad \mu_{2,n} \geq 0,$$
$$\eta_n \cdot p_{2,n} = 0, \quad \eta_n \geq 0, \quad \forall 0 \leq n \leq N. \quad (17)$$

which imply that $\lambda_{2,n}$, $\mu_{2,n}$ and $\eta_{2,n}$ are positive only when their respective constraint is active, i.e., when the battery is empty, full, or transmission power is zero respectively. As a consequence, the derivative term in (16) remains unchanged over the transmission unless one of the above events occur. The impact of each event, either increasing or decreasing the derivative term, is determined by the sign of the corresponding multiplier in (16). For the fading channel this derivative term simplifies to the sum of inverse of fading parameter $h$ and transmission power, thus yielding the well-known water-filling interpretation [16], as also seen in Subsection IV-A. However in this case, the term remains as

$$\frac{d}{dp_2}r_A(p_2) = -\frac{ap_1}{2(1 + p_1 + ap_2)(1 + ap_2)} + \frac{1}{2(1 + p_2)} \quad (18)$$

serving as the *generalized water level* to equalize throughout transmission. Therefore the same directional water-filling solution with flow constraints applies to this problem when the water level expression is replaced with the generalized level in (18). This formalization is adopted from the *generalized water-filling* approach followed in [14], and a similar solution is proposed in [10] for the multiple access channel with energy harvesters.

Since the iterative algorithm is shown to converge in Section III, an iterative algorithm alternating between directional water-filling for $T1$ and generalized directional water-filling for $T2$ converges to the optimal transmission policy for the asymmetric interference channel with $ab > 1$.

### C. Asymmetric Interference with $ab \leq 1$

In this subsection we consider the complementary asymmetric interference region to Subsection IV-B for which channel coefficients satisfy $a \leq 1$, $b \geq 1$ and $ab \leq 1$. The power-rate function for this region is given by

$$r_B(p_1, p_2) = \min \left\{ \begin{array}{l} \frac{1}{2}log\left(1 + \frac{p_1}{1+ap_2}\right) + \frac{1}{2}log\left(1+p_2\right) \\ \frac{1}{2}log\left(1 + b \cdot p_1 + p_2\right) \end{array} \right\}. \quad (19)$$

Note that the second term can also be expressed as $\frac{1}{2}log\left(1 + \frac{b \cdot p_1}{1+p_2}\right) + \frac{1}{2}log\left(1+p_2\right)$. This rate is achieved similar to the $ab \leq 1$ case by decoding the interference of $T1$ at $R2$ and treating interference as noise at $R2$. Since transmission of $T1$ needs to be decoded at both receivers in this scheme, the minimum operation decides which receiver will limit the transmission rate of $T1$.

We first tackle the single user problem for $T1$. Upon investigation, it is easily seen that the value of $p_2$ is sufficient to specify which of the two terms comes out of the minimum in (19). Specifically, the following threshold condition

$$p_2 \leq \frac{b-1}{1-ab} \triangleq p_c \quad (20)$$

implies the dominance of the first term for the sum rate in (19). Therefore, given a fixed transmission policy for the second transmitter, $\boldsymbol{p_2}$, which of the two terms will dominate the rate expression is known regardless of the value of the optimization variable $\boldsymbol{p_1}$. Therefore, the single user problem can be considered once again as a single user problem with channel fading, and can be solved through directional water-filling of [7] with the modified water base level defined as

$$\frac{1}{h} = \begin{cases} 1 + ap_2(t) & p_2(t) < p_c \\ \frac{1+p_2(t)}{b} & p_2(t) \geq p_c \end{cases}. \quad (21)$$

The single user problem for $T2$ can be solved in a similar manner to its counterpart in Subsection IV-B using a generalized water-filling algorithm with the adapted water level

$$\frac{\partial r_B}{\partial p_2} = \begin{cases} -\frac{ap_1}{2(1+p_1+ap_2)(1+ap_2)} + \frac{1}{2(1+p_2)} & p_2(t) < p_c \\ \frac{1}{2(1+bp_1+p_2)} & p_2(t) \geq p_c \end{cases} \quad (22)$$

and an alternating iterative implementation of the two single user algorithms converge to the short-term throughput maximizing power policy.

### D. Very Strong Interference

The strong interference case was identified in [17] corresponding to the case when the cross channel coefficients are large enough to ensure that the interfering signals can be decoded at both receivers. The rate for each user is therefore the single-link Gaussian channel capacity, achieving the sum rate

$$r_C(p_1, p_2) = log\left(1 + p_1\right) + log\left(1 + p_2\right), \ a > 1 + p_1, \ b > 1 + p_2 \quad (23)$$

when the conditions above are satisfied. The iterative algorithm for such a rate function is trivial; for both users, a single link short-term throughput maximization as in [5] is to be followed in each iteration. Since the two subproblems are independent, no further iterations would be necessary to reach the optimal policy. However, the nature of the problem suggests that the transmission powers of the two users are varying, and thus the requirements in (23) for the very strong interference region might not necessarily hold. The approach is justified when the output power vectors of the single user algorithms obey these constraints, which can easily be checked by comparing $a$ and $b$ with the maximum power of the two transmitters.

### E. Other Regions

In previous subsections, it is observed that some single-user subproblems reduce to modified versions of directional water-filling for asymmetric interference and very strong interference. The sum-capacity expressions for the remaining interference regions either do not have single user rate expressions yielding simpler approaches, or are not known such as the weak interference region with $\sqrt{a} + \sqrt{b} > 1$. However, we know that any two-user problem with a concave power-rate function $r(p_1, p_2)$ can be solved iteratively, using generalized directional water-filling for both single user problems. The KKT stationarity condition requires

$$\left.\frac{\mathrm{d}}{\mathrm{d}p_j}r(p_j)\right|_{p_{j,n}} - \sum_{i=n}^{N}(\lambda_{j,i} - \mu_{j,i}) = 0 \quad (24)$$

for user $j$ at the $n^{th}$ time slot, with the complementary slackness conditions

$$\lambda_{j,n}\left(\sum_{i=1}^{n}E_{j,i} - \sum_{i=1}^{n}\tau \cdot p_{j,i}\right) = 0, \quad \lambda_{j,n} \geq 0,$$

$$\mu_{j,n}\left(\sum_{i=1}^{n}\tau \cdot p_{j,i} + E_{j,max} - \sum_{i=1}^{n+1}E_{j,i}\right) = 0, \quad \mu_{j,n} \geq 0, \quad (25)$$

and the nonnegativity constraint $p_{j,n} \geq 0$. The optimal power at time slot $i$ is then $[p'_{j,i}]^+$ where $p'_{j,i}$ is the solution to (24). The interpretation of this result is the generalized directional water-filling analogy, with the alternative water level expression $\frac{\mathrm{d}r(p_j)}{\mathrm{d}p_j}$ and the same unidirectional flow and maximum flow constraints.

### V. EXTENSION TO DATA ARRIVALS

So far the focus of this paper has been on the short-term throughput maximization with only the energy causality

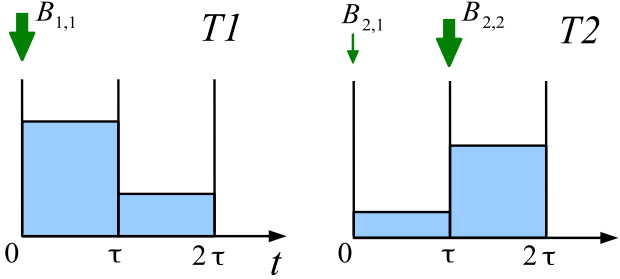

Fig. 3: A scenario where the iterative algorithm converges to a non-optimal point due to non-convex data constraints.

(2) and battery capacity (3). In this section, we provide an extension to models where the data to be transmitted is not available before transmission, but instead arrives throughout the transmission with the size of arrivals in each time slot known to the transmitters beforehand. In this setting, the transmitter aims to make the best effort to send as many of the arriving bits as possible. This problem is particularly relevant when data collection or arrival process shows a significant variation.

The problem with data arrivals is expressed in (5) with the additional data causality constraints in (5d). This set of constraints substantially affect the properties of the problem by imposing a joint constraint for the two users, as the individual rate achieved by each user is a function of both transmission powers. These constraints are not necessarily convex either, since a convex combination of two power vectors tend to increase the average achieved rate, potentially violating data causality constraints that held for the original pair of variables. Thus, the convexity and Cartesian product form properties of the constraints cease to hold, challenging the convergence of iterative algorithms for this setting.

A simple example to why a direct implementation of iterative algorithms might fail to converge to the optimal policy is presented in Figure 3. Consider two transmitters $T1$ and $T2$ with the former having a large amount of data available at the beginning of the first time slot, and the latter having only a few bits to send at first, with more data arriving at the beginning of the second time slot, as marked with the green arrows. Let the harvested energies of these two nodes be so that the water levels result as in the figure, with $T2$ sending exactly $B_{2,1}$ bits at the first time slot with this assignment. Notice that without data causality constraints, $T1$ would favor equalizing its water levels in the two slots. However, an attempt to do so in an iteration step decreases the interference of $T2$, violating its data causality constraint at the end of the first time slot. Thus an iterative algorithm is stuck at these water levels. On the other hand, if $T1$ were to equalize its water levels, $T2$ would also benefit from this since it would require less power for the first $B_{2,1}$ bits, and both users would have more power in the second time slot, yielding better performance.

As observed in the example above, the data causality constraint may cause an iterative algorithm to not converge to the global optimum. However, without this constraint, we know the convergence of the algorithm with a minimum displacement rule due to Theorem 1. Thus, in order to increase the chances of the iterative algorithm to converge to the optimal policy, we follow an alternative approach and handle the data causality constraints in a more relaxed manner to prevent cases such as in Figure 3. We suggest doing this by employing a quadratic penalty for the data constraints, the coefficient of which starts at zero and is allowed to grow indefinitely with iterations. When this approach is followed, at earlier iterations, the effect of data causality constraints is relatively small, allowing the nodes to explore the otherwise infeasible regions of the power space. However with increasing penalty coefficients, data causality constraints become strict, forcing the algorithm to converge to a power vector conforming to data-causality. The optimization problem then becomes

$$\max_{\bm{p_1} \geq 0, \bm{p_2} \geq 0} \quad \sum_{i=1}^{N} \tau \cdot r(p_{1,i}, p_{2,i}) - \epsilon_k \sum_{n=1}^{N} \|C_n\|^2 \quad (26a)$$

$$\text{s.t.} \quad \sum_{i=1}^{n} (E_{j,i} - \tau \cdot p_{j,i}) \geq 0, \quad (26b)$$

$$\sum_{i=1}^{n} (\tau \cdot p_{j,i} - E_{j,i}) + E_{j,max} - E_{j,i+1} \geq 0, \quad (26c)$$

where $C_n$ is the measure of data causality violation at the $n^{th}$ time slot, expressed as

$$C_n = min(0, \sum_{i=1}^{n} (\tau \cdot r_j(p_{j,i}) - B_{j,i})), \quad (27)$$

and $\epsilon_k$ is the penalty parameter for the $k^{th}$ iteration, increasing unboundedly with $k$.

For a continuous objective function and closed constraint set, the penalty function method is observed to have good convergence properties in practice [15], making it a good candidate for the energy harvesting problem with data constraints. The addition of the penalty term to the objective affects the water-filling algorithm for each user by creating an *offset term* in the water level expression whenever a data causality constraint is violated. This offset term is scaled by $\epsilon_k$, starts less effective and gradually becomes strict with increasing $\epsilon_k$. This can be interpreted as an additional *pump* element between time slots, forcing water-flow in either direction until the water-level difference matches the offset term. As $\epsilon_k \to \infty$, any nonzero offset term grows indefinitely, strictly requiring complying with data causality. An offset term is increasing in a users power if the corresponding constraint is of the same user, and decreasing otherwise. Thus, the pump element forces water flow in the forward direction when data causality for the user in consideration is violated, or backward when data causality for the other user is, until the difference in water level matches the offset term.

An example algorithm run is demonstrated in Figure 4 for a single transmitter with $N = 5$, energy arrivals $\bm{E} = [1, 0, 1, 0.5, 0]$, $E_{max} = 1$, data arrivals $\bm{B} = [0, 1.5, 0, 0.2, 1]$ and linear power-rate function $r(p) = \frac{p}{\tau}$ for simplicity. The taps shown in red correspond to the battery capacity constraint, and resist water flow after a total flow of $E_{max}$ including the energy harvest to the next time slot. The bidirectional pumps shown in green relate to the bit causality constraint, and activate in forward direction when more bits are being departed than received by the transmitter. Active elements of the algorithm are marked with an asterisk. The algorithm starts with the received energies placed in respective slots (b). When water flow is performed, the resulting policy and remaining active constraints are shown in (c). This corresponds to the water distribution for which any increase in water level is due to a forward pump or energy causality, and any decrease in water level is due to a reverse pump or a battery capacity tap.

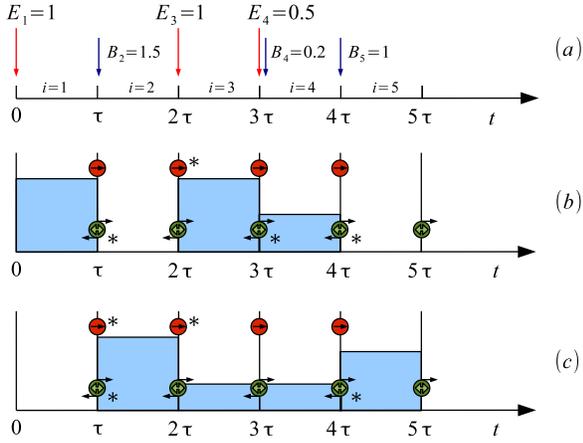

Fig. 4: (a) Energy and data arrival scenario, (b) initial water levels and (c) final water levels for a sample directional water-filling algorithm with pumps. Elements corresponding to violated or active constraints are marked with an $*$.

An interesting outcome of the pump modification is the possible tension of algorithm elements. It is possible in some extreme cases that the forward pump for data causality and the tap for battery capacity are active simultaneously. An example of this is given in Figure 5. Consider a single user problem in two time slots with $E_1 = E_2 = E_{max}$, $B_1 = 0$ and $B_2 > 0$. Since there is no data to send in the first time slot, any transmission would violate data causality and therefore the forward pump between slots 1 and 2 shown in green in Figure 5(b) stays active until there is no water in slot 1 as $\epsilon \to \infty$. On the other hand, since the second slot receives an energy of $E_2 = E_{max}$, the tap shown in red is also active, not allowing any water to be pumped into the slot. It is trivial in this example that the first energy arrival is useless and will inevitably be lost, which is what the contradiction between algorithm elements also implies. In such cases, the water causing the contradiction is removed without any reduction in the performance of the optimal policy shown in Figure 5(c).

In specific cases, the individual rate achieved by one transmitter is independent of the transmission power of the other user. Some examples to these cases are the very strong interference in Subsection IV-D, where individual rate of transmitter $j$ is $\frac{1}{2}log(1+p_j)$, or the asymmetric interference in Subsections IV-B and IV-C where the individual rate of $T2$ is $\frac{1}{2}log(1+p_2)$ regardless of the transmission power of $T1$ for $a < 1$, $b > 1$. In such cases, the cross dependence of constraints vanishes, and the directional water-filling algorithm with pump analogy is applicable without the backward-pumps. Note that in this manner, the pump extension is also a solution to the single link optimal power allocation problem with energy harvests, battery capacity and data arrivals.

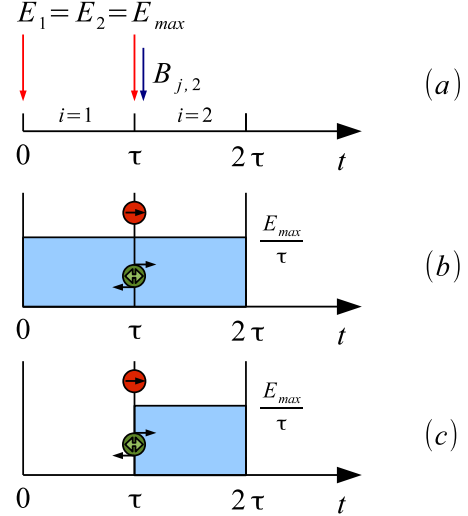

Fig. 5: Contradiction and resolution between modified directional water-filling algorithm elements.

## VI. Distributed / Online Algorithms

The optimal policies calculated using the proposed iterative algorithms require the knowledge of energy and data arrival settings of both transmitters at a centralized controller prior to the transmission to perform the iterations. In practice, such information may not be available or may not be desired to be shared. In this section, we propose near-optimal algorithms that require less information and thus are more realistic using the insight gained from the optimal iterative solution.

An important result of this paper is the convergence of single user iterative algorithm, and it is observed in Section IV that the single-user subproblems can further simplify or be independent of the other user. The role of the single user optimization problem in the optimal offline solution indicates that when the model is restricted to localized power decisions at each transmitter, i.e., without any knowledge about the energy or data arrivals of other transmitters, a reasonable algorithm is to determine policies using a single link water-filling approach while assuming expected values for the unknown parameters. In very strong interference case, this algorithm matches the optimal offline policy; whereas in weaker interference cases, further iterations only provide gradual improvements on the policy. This simplified approach performs surprisingly well, as demonstrated through simulations in Section VII.

Another case is when the energy and data arrivals are not known by the transmitters prior the transmission, and transmitters are to choose power policies as energies and packets

$$J(e_1, e_2, b_1, b_2, i) = \max_{p_1, p_2} \left( \tau \cdot r(p_1, p_2) + \mathrm{E}\left[ J\begin{pmatrix} e_1 - p_1\tau + E_{1,i+1},\ e_2 - p_2\tau + E_{2,i+1}, \\ b_1 - r_1(p_1, p_2) + B_{1,i+1}, \\ b_2 - r_2(p_1, p_2) + B_{2,i+1},\ i+1 \end{pmatrix}\right]\right) \qquad (28)$$

arrive. When data and energy harvests are independent in time, it is reasonable to represent the state of the system at any time using only Markovian states such as battery levels, data queue lengths and time before deadline. The power decision made by a transmitter at time $t$ is then a function of the subset of available states to said transmitter, depending on whether it is feasible to share the states of a node with the other. The online optimal policy in this formulation can be computationally obtained using dynamic programming techniques similar to for example in [7]. Denoting the battery and data queue states of user $j$ as $b_j$ and $e_j$ respectively, current time slot index as $i$, and assuming all states are available to all transmitters, the Bellman equation for the problem is given in (28), where $J(e_1, e_2, b_1, b_2, i)$ is the value function of a given state and E[.] is expectation based on the arrival processes. When the fixed point for the above recursive problem is found, the optimal online transmission algorithm for the interference channel becomes the arguments of the maximization in (28) for the corresponding system states.

## VII. SIMULATIONS

In this section, we present simulation results of the iterative algorithms proposed in this paper. We start by comparing the outputs of single user directional water-filling and two user iterative directional water-filling in a short time scale. We consider the throughput maximization problem in the asymmetric interference region with $ab < 1$ as in Subsection IV-B. We choose a model with time slot duration $\tau = 1\ sec$, deadline $T = 20\ sec$, battery capacities $E_{1,max} = E_{2,max} = 10\ mJ$ and channel parameters $a = 0.9$ and $b = 2$ for receiver noise spectral density $N_{0,1} = N_{0,2} = 10^{-19} W/Hz$, bandwidth $1MHz$ and direct channel coefficients $h_{11} = h_{22} = -100dB$. The sum capacity for this parameter region is given in (14) after normalization. We assume sufficient number of bits is available at both transmitters prior to transmission, and the following randomly generated energy arrival vectors

$$\boldsymbol{E}_1 = [5, 0, 0, 0, 3, 0, 0, 0, 0, 0, 7, 0, 0, 0, 4, 0, 0, 0, 6, 0]\ mJ$$

$$\boldsymbol{E}_2 = [10, 0, 7, 0, 0, 0, 0, 0, 9, 0, 0, 5, 0, 8, 0, 5, 0, 0, 0, 0]\ mJ$$

as marked with purple in Figure 6 in $mJ$ for $T1$ on left and $T2$ on right. When single user directional water-filling is performed for both transmissions independently, the resulting power policy is shown in Figure 6. However when the iterative algorithm is utilized, the optimal power policies arise as in Figure 7 shown in blue for $T1$ and red for $T2$. Recall that the algorithm for asymmetric interference channel suggests directional water-filling with base level $\frac{1}{h} = 1 + p_2$, shown in green, for $T1$; and generalized directional water-filling for $T2$. The interaction of the two power policies can be observed in the optimal policy, such as when $T1$ remains silent while $T2$ has high power at time slots $i = \{1, 2\}$, or when $T2$ significantly reduces transmission power when $T1$ is highly interfering at $i = \{19, 20\}$. The effects of such interactions is notable even in this two user model, and would be even more critical for a higher number of users sharing the same medium.

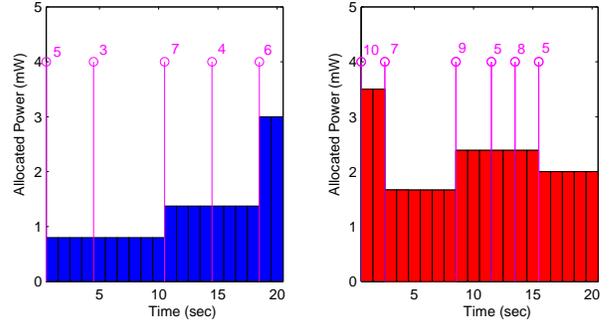

Fig. 6: Energy harvesting scenarios and power allocations with single-link directional water-filling [7] for $T1$ (left) and $T2$ (right).

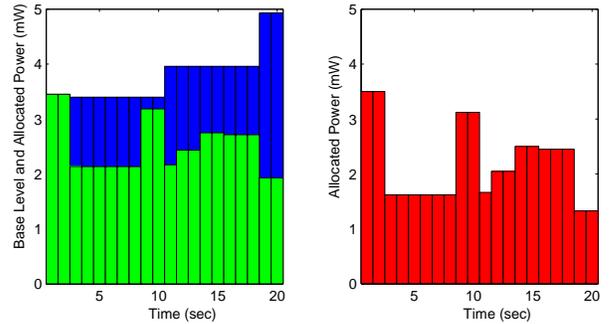

Fig. 7: Optimal power allocations with iterative directional water-filling for $T1$ (left) and $T2$ (right).

Next we compare the performances of the optimal iterative algorithm, the distributed near-optimal directional water-filling suggested in Section VI, and naive nodes that do not perform any kind of algorithm to adapt the energy harvesting process. The naive nodes attempt constant power transmission with the expected energy harvest rate at each time slot if sufficient energy is available, and transmit with all remaining energy otherwise. We assume a Gaussian interference channel with receiver noise spectral density $N_{0,1} = N_{0,2} = 10^{-19}W/Hz$, bandwidth $1MHz$ and channel coefficients $h_{11} = h_{22} = -100dB$, $h_{12} = -101.55dB$ and $h_{21} = -93.01dB$ yielding channel parameters $a = 0.7$ and $b = 5$ after normalization, falling in the asymmetric interference region of Section IV-B. For battery capacities $E_{max,1} = E_{max,2} = 10\ mJ$, we generate energy arrivals with energy distributed uniformly in $[0, E_{max}]$ and interarrival times distributed exponentially with mean $5\ sec$, quantized to time slots of duration $\tau = 1\ sec$. For this setting, the cumulative departures of these algorithms are plotted in Figure 8. It is apparent that the water-filling algorithms provide notable performance increase over the naive approach. Moreover, it is observed in this simulation as well as others with different parameters that the single user directional water-filling performs very close to optimal, making it a favorable candidate for practical applications.

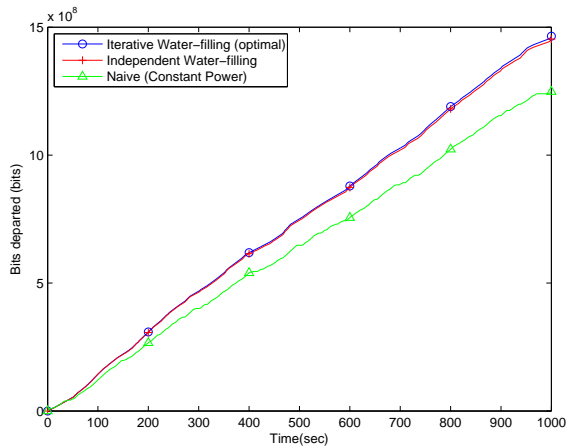

Fig. 8: Simulation of iterative directional water-filling, single user directional water-filling and naive algorithms in asymmetric interference setting with $a = 0.7$, $b = 5$ and bandwidth $1MHz$.

## VIII. Conclusion

In this paper, the short term sum-throughput maximization problem for a two-user Gaussian interference channel with energy harvesting nodes was formulated and solved with an iterative algorithm. It was observed that, in some cases, including asymmetric interference and very strong interference, the resulting generalized iterative water-filling algorithm reduces to modified versions of single-user directional water-filling. Furthermore, the model was extended to the scenario with stochastic data arrivals by introducing a pump element to the directional water-filling algorithm through a penalty for data causality violation. With the insight from the optimal solution, computationally simpler near-optimal alternatives for online and distributed versions of the problem were suggested and demonstrated along with the iterative approach. The performance of the suggested iterative directional water-filling algorithm and its distributed near-optimal counterpart were verified through simulations, showing a notable performance boost over naive algorithms.

Being the building block for multi-user interference networks, the results for the interference channel serve as a starting point for energy harvesting interference networks. This addresses practical interests in analyzing and optimizing the upcoming generation of energy harvesting networks. Future directions for this topic would be extensions to more than two users and more elaborate multi-hop network structures as well as simpler online algorithms to adapt to energy availability and interference levels simultaneously.